\documentclass[%
reprint,
superscriptaddress,
 amsmath,amssymb,
 aps,
pra,
]{revtex4-1}
\usepackage{graphicx}
\usepackage{epstopdf}
\usepackage{dcolumn}
\usepackage{bm}
\usepackage{braket}
 \usepackage{color}

 \definecolor{Revise}{rgb}{0,0,0}
\begin{document}

\preprint{APS/123-QED}

\title{A high-dimensional quantum frequency converter }
\author{Shilong Liu}
\thanks{Two authors have equal contributions on this work}
 \affiliation{Key Laboratory of Quantum Information, University of Science and Technology of China, Hefei, Anhui 230026, China; and Synergetic Innovation Center of Quantum Information and Quantum Physics, University of Science and Technology of China, Hefei, Anhui 230026, China
}
\author{Chen yang}
\thanks{Two authors have equal contributions on this work}
 \affiliation{Key Laboratory of Quantum Information, University of Science and Technology of China, Hefei, Anhui 230026, China; and Synergetic Innovation Center of Quantum Information and Quantum Physics, University of Science and Technology of China, Hefei, Anhui 230026, China
}%
\author{Zhaohuai Xu}
\affiliation{Key Laboratory of Quantum Information, University of Science and Technology of China, Hefei, Anhui 230026, China; and Synergetic Innovation Center of Quantum Information and Quantum Physics, University of Science and Technology of China, Hefei, Anhui 230026, China
}%
\author{Shikai Liu}
\affiliation{Key Laboratory of Quantum Information, University of Science and Technology of China, Hefei, Anhui 230026, China; and Synergetic Innovation Center of Quantum Information and Quantum Physics, University of Science and Technology of China, Hefei, Anhui 230026, China
}%
\author{Yan Li}
\affiliation{Key Laboratory of Quantum Information, University of Science and Technology of China, Hefei, Anhui 230026, China; and Synergetic Innovation Center of Quantum Information and Quantum Physics, University of Science and Technology of China, Hefei, Anhui 230026, China
}%
\author{Yinhai Li}
\affiliation{Key Laboratory of Quantum Information, University of Science and Technology of China, Hefei, Anhui 230026, China; and Synergetic Innovation Center of Quantum Information and Quantum Physics, University of Science and Technology of China, Hefei, Anhui 230026, China
}%
\affiliation{
Heilongjiang Provincial Key Laboratory of Quantum Regulation and Control, Wang Da-Heng Collaborative Innovation Center, Harbin University of Science and Technology, Harbin 150080, China.}
\author{Zhiyuan Zhou}%
\email{zyzhouphy@ustc.edu.cn}
\affiliation{Key Laboratory of Quantum Information, University of Science and Technology of China, Hefei, Anhui 230026, China; and Synergetic Innovation Center of Quantum Information and Quantum Physics, University of Science and Technology of China, Hefei, Anhui 230026, China
}%
\affiliation{
Heilongjiang Provincial Key Laboratory of Quantum Regulation and Control, Wang Da-Heng Collaborative Innovation Center, Harbin University of Science and Technology, Harbin 150080, China.}
\author{Guangcan Guo}
\affiliation{Key Laboratory of Quantum Information, University of Science and Technology of China, Hefei, Anhui 230026, China; and Synergetic Innovation Center of Quantum Information and Quantum Physics, University of Science and Technology of China, Hefei, Anhui 230026, China
}%
\affiliation{
Heilongjiang Provincial Key Laboratory of Quantum Regulation and Control, Wang Da-Heng Collaborative Innovation Center, Harbin University of Science and Technology, Harbin 150080, China.}
\author{Baosen Shi}
\email{drshi@ustc.edu.cn}
\affiliation{Key Laboratory of Quantum Information, University of Science and Technology of China, Hefei, Anhui 230026, China; and Synergetic Innovation Center of Quantum Information and Quantum Physics, University of Science and Technology of China, Hefei, Anhui 230026, China
}%
\affiliation{
Heilongjiang Provincial Key Laboratory of Quantum Regulation and Control, Wang Da-Heng Collaborative Innovation Center, Harbin University of Science and Technology, Harbin 150080, China.}
\date{\today}
\begin{abstract}
In high dimensional quantum communication networks, quantum frequency convertor (QFC) is indispensable as an interface in the frequency domain. For example, many QFCs have been built to link atomic memories and fiber channels. However, almost all of QFCs work in a two-dimensional space. It is still a pivotal challenge to construct a high-quality QFC for some complex quantum states, e.g., a high dimensional single-photon state that refers to a qudit. Here, we firstly propose a high-dimensional QFC for an orbital angular momentum qudit via sum frequency conversion with a flat top beam pump. As a proof-of-principle demonstration, we realize quantum frequency conversions for a qudit from infrared to visible range. Based on the qudit quantum state tomography, the fidelities of converted state are 98.29(95.02)\%, 97.42(91.74)\%, and 86.75(67.04)\% for a qudit without (with) dark counts in 2,3, and 5 dimensions, respectively. The demonstration is very promising for constructing a high capacity quantum communication network.
\end{abstract}
\pacs{Valid PACS appear here}
\maketitle
\section{Introduction}
Quantum frequency conversion enables us to change the colors of photons while maintaining its quantum properties \cite{huang1992observation}. Many quantum frequency convertors (QFCs) have been developed  based on nonlinear optical processes \cite{zhou2016orbital,steinlechner2016frequency,zhou2016orbitallight,ikuta2011wide,rakher2010quantum,tanzilli2005photonic}. For example, a polarization-insensitive QFC has been demonstrated to link an atomic ensemble and an infrared photon  \cite{ikuta2018polarization}. Nevertheless, it is still a pivotal challenge to construct a high-quality QFC for some complex quantum states, e.g., a high dimensional (HD) quantum state \cite{Yao2011a,erhard2018twisted}.

 A HD single quantum system, sometimes referred to qudit, widely existing in single atom \cite{godfrin2017operating}, photon \cite{Yao2011a,erhard2018twisted}, and superconducting quantum circuit \cite{tavakoli2015single}, has been widely applied in quantum computations and communications, for example, the Grover search algorithm \cite{godfrin2017operating,perez2018first}. In a photonic system, one of the widely used qudit is a photon state depicted by a spatial degree of freedom (DOF) i.e., orbital angular momentum (OAM) ${\left| \varphi  \right\rangle _d}{\rm{ = }}\sum\nolimits_{L = 0}^{d - 1} {{c_L}\left| L \right\rangle } $ \cite{Allen1992}, where $\ket{L}$ refers to a state with a topological charge of $L$; $c_L$($\sum c_L^2=1$) represents amplitude occupations on each eigenstate. During the last decades, many investigations have harnessed the unbounded dimensional of qudit in quantum information processing \cite{Yao2011a,erhard2018twisted,giordani2019experimental}. For example, in multilevel quantum key distribution, qudits show unique advantages not only in increasing information capacity but also in noise tolerance \cite{mirhosseini2015high,cerf2002security,Karimihd2017}. Very recently, some special fiber supporting OAM modes are developing to explore fiber-based HD-quantum communication \cite{cozzolino2019high,xavier2019quantum}, i.e., superposition OAM state \cite{cozzolino2019orbital}, and maximally entangled state \cite{cao2018distribution}.

 Frequency convertors widely exist in applications of laser source, optical communications, and some quantum protocols.  To date, almost all of FCs serve for a qubit in a two-dimensional space \cite{ikuta2011wide,rakher2010quantum,tanzilli2005photonic,ikuta2018polarization}; it is still a key challenge to construct a high-quality FC serving a qudit, for example, a FC for an OAM superposition state of $\left( {\left| 0 \right\rangle  + \left| 1 \right\rangle  + \left| 2 \right\rangle } \right)/\sqrt 3 $. The main difficulty is that the conversion efficiency (CE) of FC usually strongly depends on the topological charge $|L|$ (decreasing exponentially with $|L|$ \cite{zhou2016orbital, steinlechner2016frequency,zhou2016orbitallight,liu2017coherent,sephton2019spatial,kumar2019mode}).
This drawback prevents one from building up a high fidelity interface between photonic qudit in the different frequency domains. For example, in quantum networks with both quantum memories and fibers, most atomic based quantum memories operate in the visible wavelength \cite{duan2001long,ding2015quantum,ding2013single}, while the fiber networks connecting distributed quantum memories, usually work in the telecom band in order to minimize the transmission loss \cite{bozinovic2013terabit,wang2016advances}. Therefore, it is necessary to interface different systems in quantum communication networks by using a QFC. Many works try to balance conversion efficiency for different $L$, for example, using a short nonlinear crystal \cite{li2019frequency}, or optimizing input spatial profile \cite{kumar2019mode}; however, it reduces the overall CE.

   In this work, we first report on constructing a HD-QFC for an OAM qudit. The significant process achieved is to use a flat-top beam as a strong classical pump instead of a traditional Gaussian pump \cite{zhou2016orbital,steinlechner2016frequency,curtz2010coherent}. We get an approximate analytic expression of CE for a flat-top pump. Here, we find that the CE is insensitive to the topological charge of input states. Besides, there is no significant decrease in unit nonlinear conversion efficiency by comparing with previous various QFCs \cite{zhou2016orbital,zhou2016orbitallight,liu2017coherent,steinlechner2016frequency}. As a proof-of-principle demonstration, we realize quantum frequency conversions of a photonic qudit in dimensions of 2, 3, and 5 from infrared to visible range in low-power pump regime, where the fidelities of converted states are 98.29\%, 97.40\%, and 86.75\%, respectively. Our method would be valid for other wavelengths. The demonstrated HD-QFC fills an indispensable gap toward interference of different high dimensional systems in the frequency domain.
{\color{Revise}{
\section{Build a mode independent HD-QFC }
The second-order nonlinear process, i.e., sum-frequency generation (SFG), can be used to build a high-quality frequency convertor that connects photon state in the frequency domain. For a frequency up-conversion process, i.e., SFG in Fig. 1(a), three waves are interacting with each other in a nonlinear crystal. $E_P$ is the strong classical pump beam; $E_I$ and $E_V$ represent input signal and output photons. During this process, the energy (${\omega _P} + {\omega _I} = {\omega _V}$), the linear momentum (${k_P} + {k_I} + 2\pi /\Lambda  = {k_V}$), and the OAM (${L_P} + {L_I} = {L_V}$) are all in conservations.

   For a qudit defined in a subspace of  $\Re  \in \left\{ { - \left[ {d/2} \right],...,\left[ {d/2} \right]} \right\}$, the entire nonlinear conversion process can be described in an effective Hamiltonian \cite{zhou2016orbital,zhou2016orbitallight,kumar2019mode}:
   \begin{equation}\label{1}
      {\hat H_{eff}} = \sum\nolimits_\Re  {i\hbar {\xi _L}\left( {{{\hat a}_{I,L}}\hat a_{V,L}^\dag  + \hat a_{I,L}^\dag {{\hat a}_{V,L}}} \right)}
   \end{equation}
Where $\hat a_{I,L}^\dag$ and $\hat a_{V,L}^\dag$ represent the creating operations of infrared and visible OAM eigenstate $\ket{L}$; $\xi_{L}$ is proportional to the product of pump and the second-order susceptibility ${\chi ^{(2)}}$. During a nonlinear FC process,  the evolution of annihilation operations ($\hat a_{\left\{ {I,V} \right\},L}^\dag \left| 0 \right\rangle $) can be given in Heisenberg picture:
\begin{equation}\label{2}
   \left[ \begin{array}{l}
{{\hat a}_{I,L}}(\tau )\\
{{\hat a}_{V,L}}(\tau )
\end{array} \right] = \left[ \begin{array}{l}
\cos \left( {{\xi _L}\tau } \right)\;\; - \sin \left( {{\xi _L}\tau } \right)\\
\sin \left( {{\xi _L}\tau } \right)\;\;\;\;\;\cos \left( {{\xi _L}\tau } \right)
\end{array} \right]\left[ \begin{array}{l}
{{\hat a}_{I,L}}(0)\\
{{\hat a}_{V,L}}(0)
\end{array} \right]
\end{equation}
Here,  $\tau$ is the traveling time of the photon through the nonlinear crystal. The HD-QFC can be seen as a spatial beam splitter (BS) for OAM states in frequency domain \cite{kobayashi2016frequency}, where we can regard ${\cos ^2}\left( {{\xi _L}\tau } \right)$ and ${\sin ^2}\left( {{\xi _L}\tau } \right)$ as the transmission probability of the photon in original frequency and the reflectance probability of the photon in up-converted frequency, respectively. One should note that such a spatial BS is mode-dependent since ${\xi _L}$ is dependent on $L$.

\begin{figure}[htbp]
  \centering
  \includegraphics[width=8.5cm]{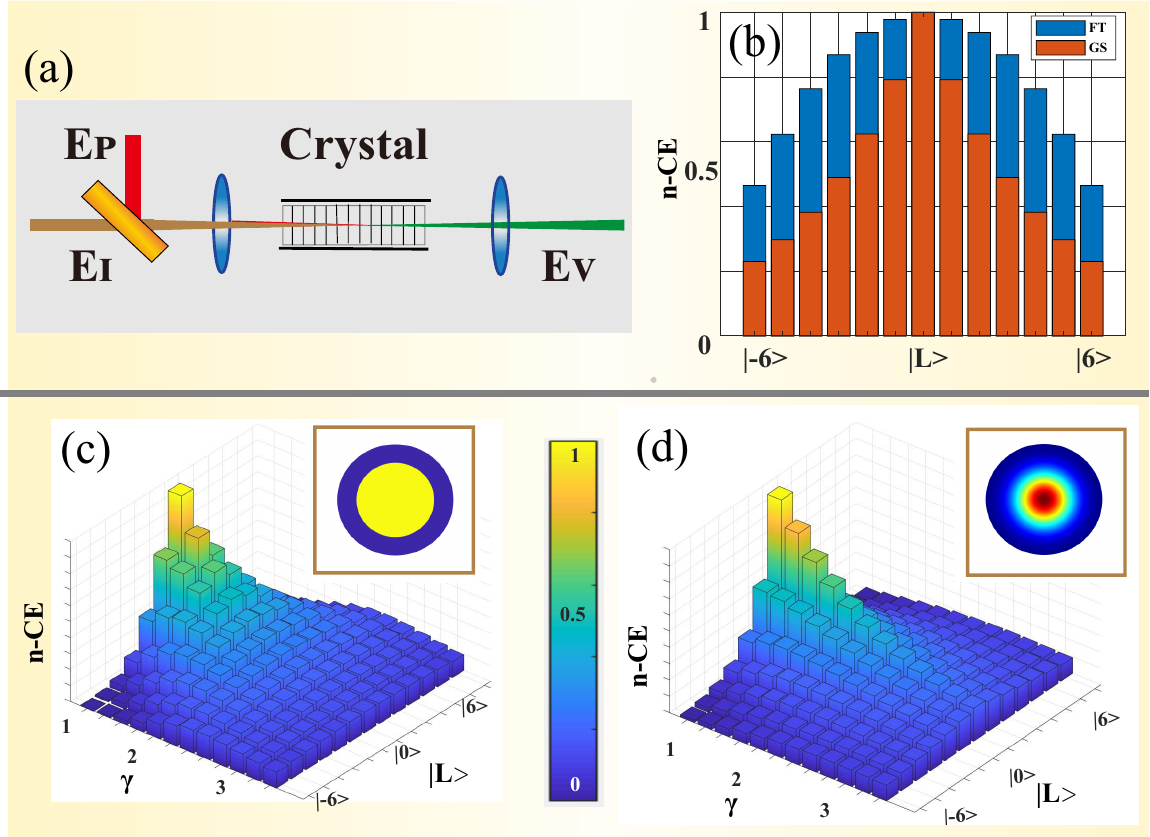}
  \caption{The conversion efficiency (CE) for HD-QFC. a: the simple regime of a sum-frequency conversion. b: the normalized CEs (n-CEs) versus different OAM laser pump of Gaussian and flat-top beam, respectively. c and d: the distributions of the c-CEs versus different beam waist ratio $\gamma$, where the inputs are Gaussian and flat-top beam, respectively. In the simulation, the wavelengths are 794 nm, 1550 nm, and 525 nm for the pump, input, and output, respectively; the beam waist is 100 um for the input signal photon; the length of the crystal(PPKTP) is 10 mm.}\label{4}
\end{figure}

For the input being a week coherent laser, one can regard  ${\cos ^2}\left( {{\xi _L}\tau } \right)$ as the conversion efficiency, which is proportional to the normalized conversion efficiency (n-CE): $\eta_P=P_V/P_IP_P $. The n-CE can be calculated via nonlinear coupled-wave equations \cite{vasilyev2012frequency,boyd2003nonlinear}.  We found that the n-CEs for different OAM laser is strongly dependent on the input pump beam profile. Generally, for a Gaussian beam, the n-CE can be given as \cite{zhou2016orbitallight,liu2017coherent}:
\begin{equation}\label{5}
  {\eta _{p- Gaussian}} = \frac{{16{\pi ^2}d_{eff}^2{L_{cry}}{2^L}}}{{{\varepsilon _0}c{n_I}{n_V}{\lambda _V^2}{\lambda _P}}}h(L,\xi )
\end{equation}
Where $L_{cry}$ is the length of crystal; $d_{eff}$ is effective coefficient of the crystal; $n_{V,I}$ belongs to the refractive index for up-converted and input fields; $\varepsilon_{0}$ and $c$ represent the vacuum permittivity and the light speed in the vacuum; $h(L,\xi)$ is an integral function associating with focusing parameter $\xi=L_{cry}/2\pi w_0^2/\lambda_P$ and topologic charge $L$. The n-CES for OAM state from -6 to 6 are shown in Fig. 1(b) (red bars), where beam waists for input and pump  set 100 um. We can find that the n-CEs will decrease rapidly along with the increases of $L$. However, the n-CEs will have a considerable improvement if the input is a flat-top beam.
\begin{figure*}[htb]
  \centering
  \includegraphics[width=16cm]{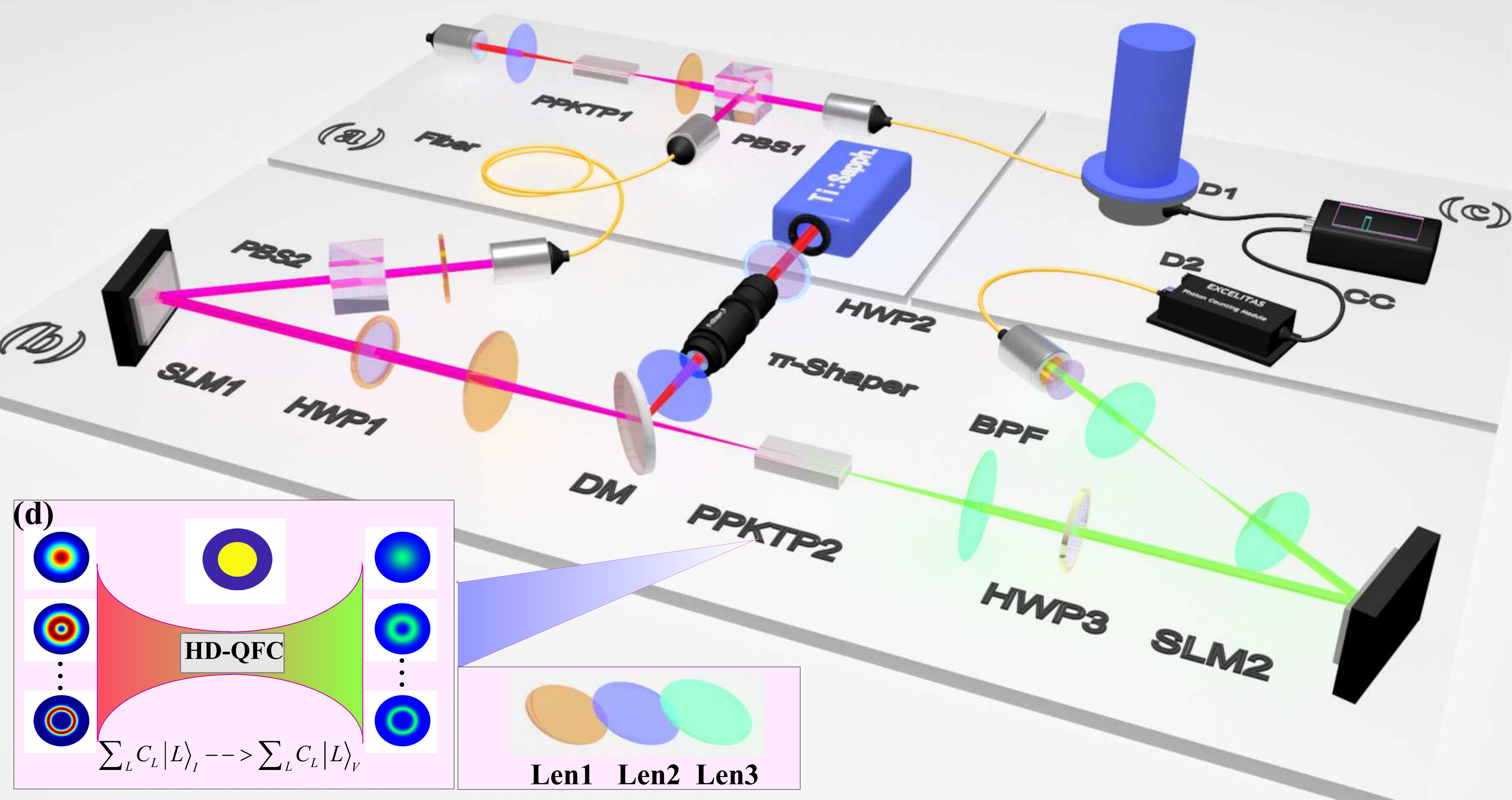}
  \caption{Schematics of a HD-QFC. a: Generation of an infrared heralding single-photon state. b: the HD-QFC for a qudit preparation, conversion, and projection measurement, where the pump is a flat-top beam shaped by a  $\pi$-shaper. c: The electronic logic modules for correction measurement. d: The simple frame for the HD-QFC. PBS: polarization beam splitter; DM: dichromatic mirror; HWP1 (2,3): half-wave plate for infrared (pump, visible) photon; Len1(2,3): lens work in infrared (1550 nm), pump (794 nm), and visible (525 nm) bands. SLM1(2): a spatial light modulator for infrared (visible) photons; BPF: bandpass filter; D1: D2: single-photon avalanche diode detector; C.C.: correction coincidence logic. The classical strong pumps in SPDC and sum frequency generation are continuous waves.}\label{4}
\end{figure*}
The FTB has a flat intensity profile and flat phase ($L=0$) in some well-defined region, and zero elsewhere:
  \begin{equation}\label{4}
  {I_{FTB}}(\rho ) = \left\{ \begin{array}{l}
1\;\;\;\;\;\;\;\;\left| \rho  \right| \le {w_{FTB}}\\
0\;\;\;\;\;\;\;\;\left| \rho  \right| > {w_{FTB}}
\end{array} \right.
  \end{equation}

Where $w_{FTB}$ is the width of the FTB; $\rho$ (=$\sqrt{x^2+y^2}$) is the spatial transverse coordinates.  By inserting the flat-top beam into the coupling equation, one can get the n-CEs for an OAM eigenstate:
\begin{equation}\label{5}
  {\eta _{p-flattop}} =\frac{{16\pi d_{eff}^2{2^L}}}{{{\varepsilon _0}c{n_V}{n_I}{n_p}\lambda _P^2w_p^2L!}}h(L,\gamma )
\end{equation}
Where the $h(L,\gamma)$ is an integral function associating with beam waists ratio ($\gamma=w_p/w_i$) and topologic $L$. We make a calculation of the n-CEs under the flat-top beam, which is shown in Fig. 1(b) (blue bars). Here, all the parameters are the same as the Gaussian's situation. The simulations illustrate that the n-CEs under the flat-top pump tend to be flatter than the Gaussian pump, which is beneficial to build a mode-dependent QFC in an OAM subspace. Also, we studied the n-CEs again the different beam ratio of $\gamma$, which is shown in Fig. 1(c)(flat-top) and Fig. 1(d)(Gaussian). Two 3-D distributions of n-CEs clearly illustrate there is a considerable improvement in the regime of the flat-top beam. The larger beam size ($\gamma$) of the flat-top beam has, the more flat n-CEs of the OAM state will.  It is reasonable because the flat-top beam is an ideal plane wave when the beam size is large enough.

The n-CE in Eq. (5) is the ideal situation that the beam profile still keeps a flat in crystal. However, the flat-top beam will be out of shape due to the diffraction, which is because the flat-top beam is not the solution of the paraxial Helmholtz equation. One could first determine $E_P(\rho,z)$ by Rayleigh Sommerfeld diffraction integral \cite{gillen2016comparison,brosseau1998fundamentals}. Then, one can numerically simulate the output field $E_V$ via coupling equations. Fortunately, the flat-top beam shaped by $\pi$-shaper still keeps a flat profile within 10 mm. Therefore, the approximation is reasonable. (The more details can be seen in appendix A)}}
\section{Results}
\subsection{Experimental setup for a HD-QFC}
 Fig. 2 shows an experimental setup of the HD-QFC, which can converts a qudit from infrared to the visible spectrum. Three parts are assigned to state generation (Fig. 2a), conversion (Fig. 2(b)), and correction measurements (Fig. 2c), respectively. The infrared heralded single photons are prepared via the spontaneous parametric down-conversion (SPDC) in a type-II quasi-phase-matching nonlinear crystal (PPKTP1, $1\times2\times20$ mm$^3$).
 The waist of the Gaussian beam with 775 nm is 115 um at the center of PPKTP1. The infrared 1550 nm photon pairs are collected by an infrared lens (f=100 mm). The infrared idler photon acts as a heralding signal detected by a superconducting nanowire single-photon detector (SNSPD); the signal photon is collected to the fiber as a source to prepare an arbitrary OAM qudit by a spatial light modulator (SLM1). The heralded infrared single-photon efficiency is 25\%. For reducing the random coincidence, we add a narrow band fiber filter of 100 GHz bandwidth at the center of 1550 nm.

 For qudit state generation, one can employ amplitude-encoding technology with the help of a SLM \cite{bolduc2013exact}. In our recent work \cite{liu2019classical}, we can prepare an arbitrary OAM superposition state $\sum_{L}C_L\ket{L}$ with high fidelity at listed in the dimension of 7. Following that technology, we first create the phase hologram of a qudit, then load the phase hologram onto SLM2. The quality of the state can be tested by interference visibility or fidelity with the projection measurements. One thing that we need to is to slightly balance the amplitudes for each OAM eigenstate to get rid of the mode-dependent reflective efficiency.

 The HD-QFC (Fig. 2(d)) plays the role of a coherent interface connecting the infrared and visible photon via sum frequency generation (SFG),
 SFG involves a with a flat-top profile ($E_P$, $L$=0), and a type-I quasi-phase-matching nonlinear crystal (PPKTP2). Generally, the CEs for different OAM eigenstates decrease rapidly along with an increase of $L$ \cite{zhou2016orbital,liu2017coherent,sephton2019spatial} when the pump is a Gaussian beam. As the theoretical predictions, this drawback can be overcome if one uses a flat-top beam (FTB) as a pump. In our setup, both $\pi$-Shaper and a Fourier len are used to transform a Gaussian beam to be a focused flat-top beam. The basic principle of $\pi$-Shaper is to shape a Gaussian beam to an Airy disk via the Fourier-Bessel transform: ${I_f}(\rho ) = {I_{f0}}{\left[ {{J_0}(2\pi \rho )/2\pi \rho } \right]^2}$ \cite{laskin2013beam}. Here ${J_0}(2\pi \rho )$ is the first kind and zero-order Bessel function; $I_{f0}$ is the normalized factor. The beam profile of the flat-top can be seen in appendix A.

  \subsection{Conversion efficiency and coherence of qudits}
    Fig. 3(a) shows the relationship between the power CE against the input pump power for OAM eigenstates $\ket{0}_I$, $\ket{1}_I$, and $\ket{2}_I$, where the insets are corresponding theoretical intensity distributions. In Fig. 3(a), the power CE is proportional to the input pump's power. Therefore, one can calculate the average normalized nonlinear CE (n-CE) (${\rm{ = }}{P_V}{\rm{/}}{P_I}{P_P}$): 0.37\%/W, 0.42\%/W, and 0.33\%/W for the OAM eigenstate of $L$=0, 1, and 2, respectively. On the contrary, the previous schemes with a Gaussian pump show big different n-CE among these eigenstates \cite{zhou2016orbital,liu2017coherent,kumar2019mode} (also see the Table 1 in appendix A), for example, 1.5\%/W, 0.5\%/W, and 0.3\%/W for the same states, respectively \cite{liu2017coherent}. The theoretical n-CE for two types of pumps can be calculated by the coupled-wave equations \cite{vasilyev2012frequency} (also see appendix A). Both theory and experiments show the n-CE are nearly equal for three eigenstates in our scheme, which enable us to build a high-quality HD-QFC in a five-dimensional subspace.

Fig. 3(b) depicts the coincidences between the input infrared and converted visible photons, where the topologic charges of the input and projected states set to be -3 to 3. For quantifying crosstalk, we calculate the visibility (=$\sum\nolimits_i {{C_{ii}}/} \sum\nolimits_{i,j} {{C_{i,j}}} $) being 90.53(83.47)$\pm$ 0.68\% without (with) dark counts, where the errors are ±1 standard deviations (std) assuming the data follows Poisson's distribution. Due to the mode-dependent collection efficiency in the projection measurement, the coincidence for higher-order mode decreases quickly. Nevertheless, one can construct a HD-QFC for a qudit consisting of several symmetric low order modes, for example, a five-dimensional qudit.

For a qudit defined in a subspace $\Re $, an ideal output state after HD-QFC can be written as:
\begin{equation}\label{3}
{\left| \varphi  \right\rangle _\Re }{\rm{ = }}\frac{1}{{\sqrt d }}\sum\nolimits_{j =  - [d/2]}^{[d/2]} {\left| j \right\rangle }
\end{equation}

Usually, there are some relative phases  ${e^{i{\phi _L}}}$ between different OAM eigenstates due to mode dispersion \cite{liu2018coherent}. For testing the existing phases, we perform projection measurements in SLM2 (in Fig. 2) via scanning one of the phases to get coincidence interference curves, which should be given by  ${\left| {d - 1 + {e^{i{\phi _L}}}} \right|^2}$ for an ideal qudit.

  For a two-dimensional state, the theoretical coincidence should be $1 + \cos \left( {{\phi _L}} \right)$; the visibility (${\rm{ = }}{C_{Max}} - {C_{Min}}/{C_{Max}} + {C_{Min}}$) could be up to 100\%. Fig. 3(c) shows the experimental data for a two-dimensional state $\left( {\left| { - 1} \right\rangle  + \left| 1 \right\rangle } \right)/\sqrt 2 $, where each of data is recorded per 60 s without the dark counts. The experimental visibility is 95.32$\pm$2.60\% according to the solid fitted line, where the error bars represent 2 stds assuming coincidences follows Poisson's distribution. Fig. 3(d) is the situation of a three-dimensional qudit, where the visibility is 75.97$\pm$ 4.58\%. For a three-dimensional qudit, the theoretical visibility is 80\%. Because of a basic phase between different modes \cite{liu2018coherent}, the fitted line appears a very small shift. During the state preparation, we can make up for this small phase by adjusting the input phase hologram in SLM1. Nevertheless, the visibility illustrate the qudit states have good coherence.

\subsection{Quantum state tomography of qudits}

We evaluate a photonic qudit in subspace $\Re$ via the qudit quantum state tomography (QST) \cite{thew2002qudit} after HD-QFC. In order to do that, we need to make a mode projection-measurement in the complete mutually unbiased bases (MUBs). For a qudit defined in the prime dimension, the used MUBs can be generated by a discrete Fourier transformation \cite{wiesniak2011entanglement}:
\begin{equation}\label{2}
  \left\{ {{\rm{|a}}_m^j{\rm{ > }}} \right\} = \left\{ {\frac{1}{{\sqrt d }}\sum\limits_{n = 0}^{d - 1} {\omega _d^{\left( {j{n^2} + nm} \right)}} \left| n \right\rangle } \right\}
\end{equation}

 Where $j$ indexes the group of the MUBs; $m$ indexes the superposed OAM states for each sets \cite{wiesniak2011entanglement,liu2019classical} (also see appendix B).
 After loading the phase hologram of MUBs (Eq. 7) in SLM2, we perform correction measurements between infrared and visible photons. Using the maximum-likelihood estimation method, we reconstruct the density matrix of a qudit, which is shown in Fig. 4. Fig. 4 (a) and (b) are the real and imaginary parts of the density matrix of the qubit state, where we perform QST in two-dimensional subspace $\ket{\varphi}_{\Re =2}= {\left| {-1} \right\rangle  + \left| 1 \right\rangle } /\sqrt 2$. Fig. 4(c)-(d) and Fig. 4(e)-(f) are the situations of ${\left| \varphi  \right\rangle _{\Re {\rm{ = }}3}}$ and ${\left| \varphi  \right\rangle _{\Re {\rm{ = 5}}}}$  in their subspaces.
\begin{figure}[htbp]
  \centering
  \includegraphics[width=8cm]{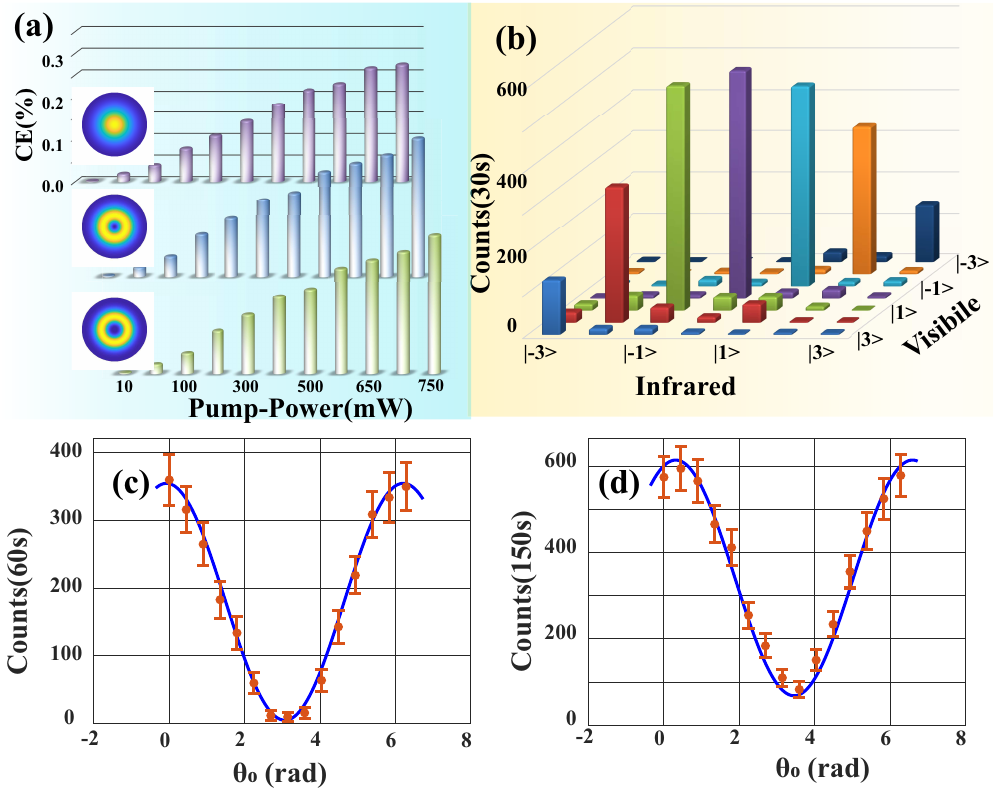}
  \caption{The mode conversion efficiency of HD-QFC and interference curves for the qudit in the dimension of 2 and 3. a: The mode conversion efficiencies versus the input pump power for inputs of  $\ket{0}$, $\ket{1}$, and $\ket{2} $. b: The coincidence between infrared and visible photons for single OAM eigenstates in subspace $\left\{ { - 3,...,3} \right\}$, where the dark count is 6. c and d :Coincidences subtracted dark counts are recorded by scanning the phase angle in spatial light modulator (SLM2), where the projection-states are  ${\left( {{e^{i{\theta _0}}}\left| -1 \right\rangle  + \left| 1 \right\rangle } \right)^\dag }$ and ${\left( {\left| { - 1} \right\rangle  + {e^{i{\theta _0}}}\left| 0 \right\rangle  + \left| 1 \right\rangle } \right)^\dag }$ for a two and a three dimensional qudit state, respectively. Each coincidence data for b, c, and d are recorded by 30s, 60s, and 150s. }\label{2}
\end{figure}

\begin{figure}[htb]
  \centering
  \includegraphics[width=8cm]{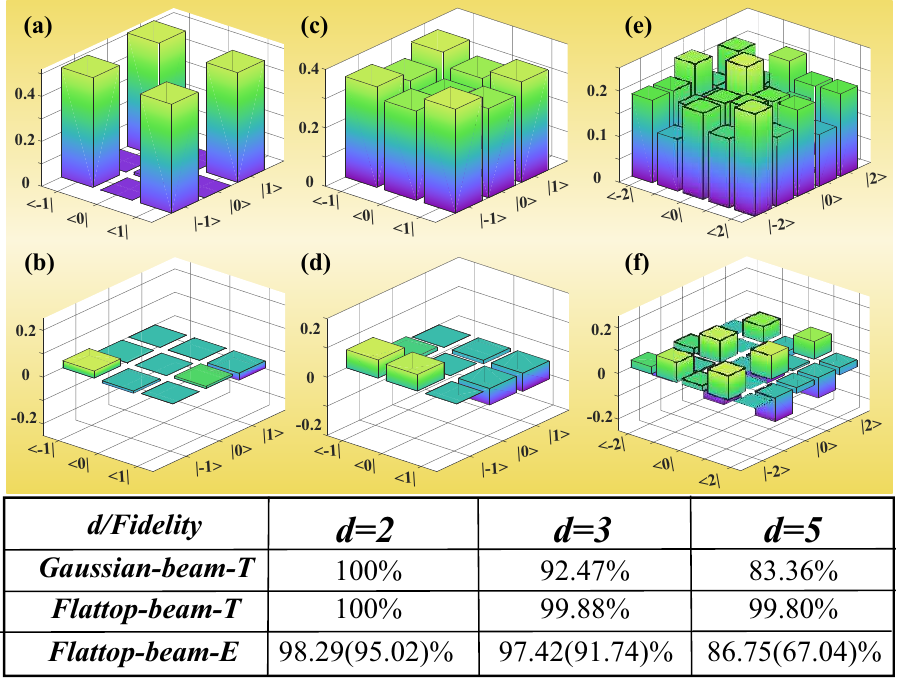}
  \caption{The reconstructed density matrix and the fidelity for a qudit in three- and five- dimensional space via quantum state tomography. a, b: The real and imaginary parts of the density matrix for a qubit state $\left( {\left| { - 1} \right\rangle  + \left| 1 \right\rangle } \right)/\sqrt 2 $. c, d: The real and imaginary part of the density matrix for a three-dimensional state  ${\left| \varphi  \right\rangle _{\Re {\rm{ = }}3}}\left( {\left| { - 1} \right\rangle  + \left| 0 \right\rangle  + \left| 1 \right\rangle } \right)/\sqrt 3 $ . e, f: are the situations of a qudit ${\left| \varphi  \right\rangle _{\Re {\rm{ = }}5}}{\rm{ = }}\left( {\left| { - 2} \right\rangle  + \left| { - 1} \right\rangle  + \left| 0 \right\rangle  + \left| 1 \right\rangle  + \left| 2 \right\rangle } \right)/\sqrt 5 $ in five-dimensional subspace. Each coincidence data for three qudits are recorded at 100s, 300s, and 300s, respectively. The corresponding dark counts are 16, 20 and 40, where the single counts in idler ports are 0.4MHz,0.4MHz, and 1Mhz. The inserted table shows the fidelities for qudit in d=2, 3 and, 5 via Gaussian or flat-top beam pump, where the Gaussian-beam-T and Flat-top-beam-T are theoretical predictions based on the nonlinear coupling equations; the Flattop-beam-E is the experimental reconstructed results based on the state tomography. }\label{3}
\end{figure}

  Usually, we tend to calculate the fidelity by $F = Tr{\left[ {\sqrt {\sqrt \rho  {\rho _{\exp }}\sqrt \rho  } } \right]^2}$  between theoretical ($\rho {\rm{ = }}{\left| \varphi  \right\rangle _\Re }{\left\langle \varphi  \right|_\Re }{\rm{ = }}1/d \cdot \sum\nolimits_{j,k =  - [d/2]}^{[d/2]} {\left| j \right\rangle \left\langle k \right|}$) and experimental density matrixes to evaluate the quality of the QFC. The fidelities without (with) dark counts are 98.29(95.02)$\pm$1.55\%, 97.42(91.74)$\pm$1.11\%, and 86.75(67.04)$\pm$1.80\% for a two-, three-, and five-dimensional qudit. The theoretical fidelities of QFC with Gaussian and a flat-top beam as a pump are summarized in a table below Fig. 4.

{\color{Revise}
The fidelities are a bit low for a qudit in a five-dimensional subspace because of the low signal-noise ratio (SNR). Also, the SNR will put the visibility down during the interference measurements. SNR comes from two-part: the accidental counts and the signal coincidence counts. Generally, the less accidental counts or, the more signal coincidence counts, then the higher SNR, will produce the higher fidelity of qudit. The accidental counts mainly come from background electric noise, stray light, and spontaneous radiation noise, which approaches a constant when input parameters are fixed. However, the mode signal coincidence counts mainly depend on the conversion and collection efficiency. In our regime, the conversion efficiency is equally at last in a five-dimensional subspace $\{-2,-1,...,2\}$, while the effective collection efficiency is strongly dependent on the dimensions if one uses the mode projection-measurement \cite{poyatos1997complete,bouchard2018measuring}. For example, for a two dimensional state, $(\ket{-1}+\ket{1})/\sqrt{2}$, the state will be $(\ket{-2}+\ket{0}+\ket{0}+\ket{2})/4$, where we only collect the $\ket{0}$ due to a single-mode filter. Therefore, only 50\% photons are collected. For a qudit in $d$ subspace, the effective coupling photon occupies only $1/d$. So in our experiment, the fidelity of qudit in a five-dimensional subspace mainly comes from the natural dimension-dependent collection efficiency during the projection- measurement. Therefore, the subtracting the accidental counts is reasonable during the state constructions, especially for a qudit in higher dimensions.  Besides, some minor factors could affect the fidelity during the QST, i.e., the imperfect input state, mode-dependent losses (transmission) \cite{krenn2015twisted}, and mode dependent collection \cite{qassim2014limitations}. Nevertheless, we give two results in visibility and fidelity with and without dark counts}

\section{Discussion}
Employing the pump manipulating technology, we firstly build a high-quality HD-QFC connecting two photons in different colors. Using a focused flat-top beam as a pump not only keeps the input beam profiles but also has a normal normalized conversion efficiency. In the future, we can build a higher dimensional QFC to serve for high dimensional quantum communications. Based on the Eq. (5), we need to optimize the area of the flat-top beam in the nonlinear crystal. Also, the HD-QFC with a flat-top beam pump maybe presents a unique advantage in image frequency conversion as it supports the more high-order spatial modes. The technique could be beneficial to other conversion systems, i.e., the reversible OAM photon–phonon conversion \cite{zhu2016reversible}.

Because the current QFC works in a single pass regime, the overall CE is rather low. Nevertheless, one can find that the normalized mode conversion efficiency is comparable with other implementations (see Table1 in appendix A). The total CEs would increase via directly enhancing input pump laser power or placing the crystal in a particular cavity that resonates with a flat-top beam \cite{naidoo2018brightness}. Recently, some flat-top beam-shaping technique and productions have been proposed to work in higher power \cite{bovatsek2007high, laskin2016refractive}. Another convenient way is to employ a high peak power pulse as a classical pump, where the photon pairs should also be generated from a pulse laser pump.

\section{Acknowledgments}
This work is partially supported by the Anhui Initiative in Quantum Information Technologies (AHY020200); National
Natural Science Foundation of China (61435011, 61525504, 61605194, 61775025, 61405030,11934013); China Postdoctoral Science Foundation (2016M590570, 2017M622003); Fundamental Research Funds for the Central Universities.
\appendix
\section{The conversion efficiency under the pumps of Gaussian and flat-top beam}

 For a frequency up-conversion process, three waves are interacting with each other in a nonlinear crystal. When the pump has strong power, the frequency up-conversion process can be calculated by the nonlinear coupled-wave equations \cite{vasilyev2012frequency}:
  \begin{equation}\label{1}
\begin{array}{l}
\frac{{\partial {E_I}(\vec \rho ,z)}}{{\partial z}} = \frac{i}{{2{k_I}}}{\nabla _\rho }^2{E_I}(\vec \rho ,z) + {K_I}E_P^*(\vec \rho ,z){E_V}(\vec \rho ,z){e^{i\Delta kz}}\;\;\\
\frac{{\partial {E_V}(\vec \rho ,z)}}{{\partial z}} = \frac{i}{{2{k_V}}}{\nabla _\rho }^2{E_V}(\vec \rho ,z) + {K_V}{E_P}(\vec \rho ,z){E_I}(\vec \rho ,z){e^{ - i\Delta kz}}
\end{array}
  \end{equation}
Where the pump $E_P$, input signal $E_I$, and the up-converted output $E_V$ have the angular frequency of $\omega _P$, $\omega_I$, and $\omega_V$, respectively; $k_I$ and $k_V$ represent the wave vector of the input infrared and output visible fields. $\nabla _\rho ^2{\rm{ = }}{\partial ^2}/\partial {x^2} + {\partial ^2}/\partial {y^2}$ is two dimensional Laplace operator; $\Delta {k} = {k_p} + {k_I} - {k_V} - 2\pi /\Lambda $ show the phase mismatch equation in wave vector domain. One can get the output field $E_V(\rho,z)$ via solving coupling Eq. A1 with the split-step Fourier method. In that case, the output power can be calculated after the nonlinear conversion:
\begin{equation}\label{2}
  {P_V}(z = {L_{half}}) \sim \int\limits_{ - {L_{half}}}^{ + {L_{half}}} {{E_V}(z)} E_V^*(z)dz
\end{equation}
Where $L_{half}$ is the half length of the nonlinear crystal ($L_{cry}$). Based on these equations, the normalized power conversion efficiency(n-CE) can be written as $\eta_p=P_v/P_IP_P$, and the quantum situation has the such form: ${\eta _q} = {N_V}/{N_I} = {P_V}{\lambda _V}/{P_I}{\lambda _I}$. Where $\lambda_I$ and $\lambda_V$ are wavelengths of input and output photons, respectively. We now discuss two types of up-conversion process based on the pump beam profile.

i) The pump is the Gaussian beam. Based on the nonlinear coupling equation, one can get the analytical expression of the n-CE as the input being a Laguerre-Gaussian(LG) mode \cite{zhou2016orbitallight,liu2017coherent}:
\begin{equation}\label{5}
  {\eta _{p- Gaussian}} = \frac{{16{\pi ^2}d_{eff}^2{L_{cry}}{2^L}}}{{{\varepsilon _0}c{n_I}{n_V}{\lambda _V^2}{\lambda _P}}}h(L,\xi )
\end{equation}
Where $L_{cry}$ is the length of crystal; $d_{eff}$ is effective coefficient; $n_{V,I}$ belongs to the refractive index for up-converted and input fields; $\varepsilon_{0}$ and $c$ represent the vacuum permittivity and the light speed in the vacuum; $h(L,\xi)$ is an integral function associating with focusing parameter $\xi=L_{cry}/2\pi w_0^2/\lambda_P$ and topologic charge $L$.

ii) The pump is a flat-top beam. A flat-top beam. i.e.,
\begin{equation}\label{8}
  {E_p} = \left\{ \begin{array}{l}
{N_0}\;\;r <  = {w_p}\\
0\;\;\;\;r > {w_p}
\end{array} \right.\;\;
\end{equation}
is a flat intensity in a special area ($r<=w_p$), and zero intensity in otherwise ($r>0$). We can get the normalized $N_0$ based on the input pump power: ${N_0} = \sqrt {{P_P}/2\pi {\varepsilon _0}c{n_p}w_p^2} $. We can calculate the n-CE via coupling equation when the pump is strong laser:
\begin{equation}\label{9}
  {\eta _{p- flattop}} = \frac{{16\pi d_{eff}^2{2^L}}}{{{\varepsilon _0}c{n_V}{n_I}{n_p}\lambda _P^2w_p^2L!}}h(L,\gamma )
\end{equation}
Where all of the parameters are same as the situation of Gaussian pump. $h(L,\gamma)$ is the integral function with topologic $L$ and beam waist ratio $\gamma(=w_p/w_i)$. It can be given in:
\begin{equation}\label{10}
\begin{array}{l}
h(L,\gamma ) = \int\limits_{ - {L_{cry}}/2}^{{L_{cry}}/2} {\int\limits_{ - {L_{cry}}/2}^{{L_{cry}}/2} {\left( {L! - \Gamma (1 + L,b{\gamma ^2}))} \right)} } \\
\;\;\;\;\;\;\;\;\;\;\;\;\;\;\;\;\;\;\;\; \times \left( {1 + \frac{{ix}}{{{Z_I}}} + 1 - \frac{{iy}}{{{Z_I}}}} \right)dxdy
\end{array}
\end{equation}
Where $Z_I(=\pi w_i^2n_I/\lambda_I)$ is the Rayleigh distance of the input infrared light; $\Gamma (n,z)$ represents the incomplete gamma function. Based on the Eq. A5, one can get the n-CEs for different OAM states. Also, one needs to note the topologic $L$  should be more than -1 in Eq. A5. Therefore, the n-CE for the negative OAM state is calculated by the corresponding positive OAM state.

 Because the flat-top beam is not the solution of the paraxial Helmholtz equation, the flat-top beam will be out of shape due to the diffraction. In order to get the output $E_V$, one can first determine $E_P(\rho,z)$ by Rayleigh–Sommerfeld diffraction integral \cite{gillen2016comparison,brosseau1998fundamentals}. Then, the output field $E_V$ and original $E_I$ can be simulated via Eq. A1.
 In that case, we can calculate all of the parameters, i.e., conversion efficiency, state fidelity, and so on.

\begin{figure*}
  \centering
  \includegraphics[width=16cm]{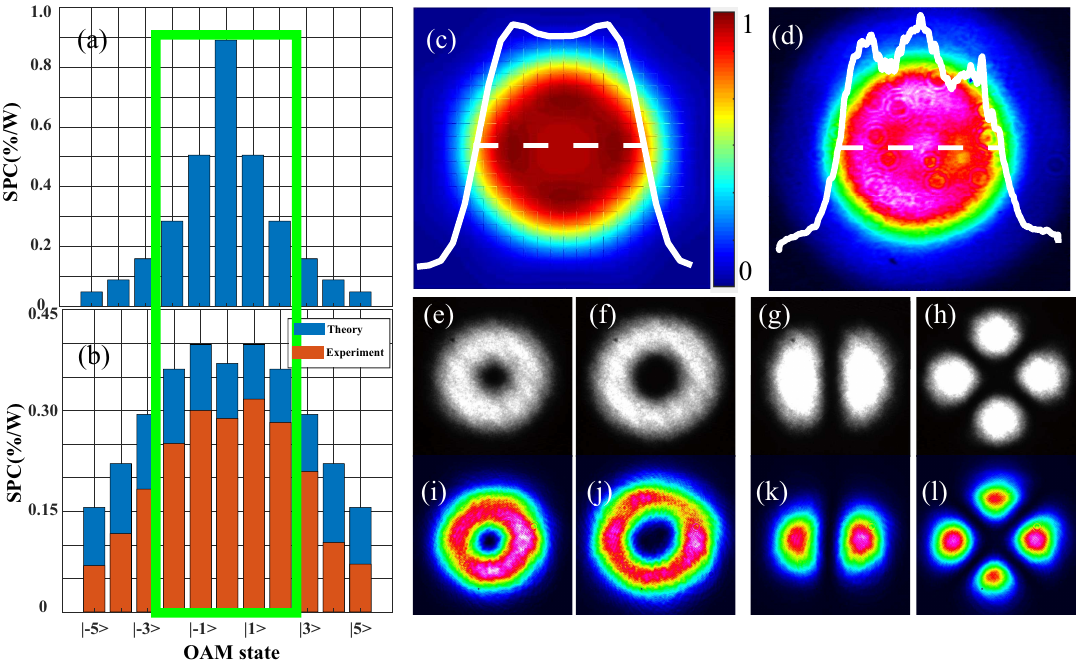}
  \caption{The conversion efficiency and beam profiles for a high-dimensional frequency converter (HD-FC). (a)-(b): Conversion efficiency for the Gaussian and flat-top beam pump in a single-pass configuration. (c)-(d): The beam profiles for theoretical and experimental flat-top beam, where the white lines are the one-dimensional distribution along the vertical center of the beam profile. (c)-(h): The beam profiles of input OAM states. (i)-(l): The beam profiles of up-converted OAM states after HD-FC. }\label{1}
\end{figure*}
   Fig. 5(a) and Fig. 5(b) are the theoretical single-pass conversion efficiency (SPCE) when the input pump is a Gaussian and a flat-top beam, respectively. We use the split-step Fourier method to solve the Eq. A1, thus get the corresponding power conversion efficiency. In our scheme, we employ a  $\pi$- shaper to generate an Airy disk, then use a Fourier lens to transfer an Airy disk to a focusing flat-top beam in the Fourier plane \cite{laskin2013beam}. Here the diameter of the pump beam before  $\pi$- shaper is 4 mm, the distance between the Fourier lens (f=300 mm)  and  $\pi$- shaper is 100 mm. The length of the crystal (PPKTP) is 10 mm. We simulate the distribution of the flat-top beam in the center of crystal via the Collin diffraction equation \cite{collins1970lens}, which is shown in Fig. 5(c). The red bars in Fig. 5(b) shows the experimental SPCE with the flat-top pump. We can find that SPCEs are flat for several low OAM modes, i.e., L=-2 to 2, which are marked on the green box. However, in the situation of the Gaussian beam pump, there is a big gap in SPCEs among those OAM modes. Table1 shows CEs of various frequency convertors for single-photon or entangled states. We can find that the n-CEs converted by flat-top beam near equally for different OAM states in five-dimensional space.

Fig. 5(d) is the intensity of a flat-top beam profile in the center of crystal acquired by a CCD. Because of the imperfect alignment, the flat-top beam is not ideal. Nevertheless, the experimental results show unique advantages in OAM mode frequency conversion. Fig. 5(e)-(l) show the input (infrared) and output (visible) beam profiles of a single and superposed OAM states, respectively. We prepare the infrared OAM eigenstates of  $\ket{1}$, $\ket{2}$, $\ket{-1}+\ket{1}$, and $\ket{-2}+\ket{2}$ , respectively, which is shown in Fig. 5(e)-(h). Fig. 5(i)-(l) show the corresponding visible beam profiles. The high similarity in the beam profile can directly illustrate that the reliability of HF-FC.

\begin{table*}[]
\center
\begin{tabular}{|c|c|c|c|c|}
\hline
Dimensions& Configuration&Converted state & N-CE(\%/W)& Ref.\\
\hline
d=2
& \begin{tabular}{c}
    Cavity, 10mm PPKTP \\
    (1560nm+792nm=525nm)\\
    \end{tabular}
&\begin{tabular}{c}
   Single photon state:\\
   $\ket{L}+e^{i\phi}\ket{-L}$ \\
 \end{tabular}
&  \begin{tabular}{c}
 3.5@L=0\\
1.4@L=1\\
0.4@L=2\\
    \end{tabular}
&\cite{zhou2016orbitallight}\\
\hline
d=2
& \begin{tabular}{c}
    Cavity, 10mm PPKTP \\
    (1560nm+792nm=525nm)\\
    \end{tabular}
&\begin{tabular}{c}
   Entangled state:\\
   $\ket{H,1}+e^{i\phi}\ket{V,-1}$ \\
 \end{tabular}
&  \begin{tabular}{c}
1.0@L=1
    \end{tabular}
&\cite{zhou2016orbital}\\
\hline
d=2
& \begin{tabular}{c}
    Cavity, 50mm PPLN \\
    (1560nm+792nm=525nm)\\
    \end{tabular}
&\begin{tabular}{c}
   Coherent laser:\\
   $\ket{L}+e^{i\phi}\ket{-L}$ \\
 \end{tabular}
&  \begin{tabular}{c}
 1.5@L=0\\
0.5@L=1\\
0.3@L=2\\
    \end{tabular}
&\cite{liu2017coherent}\\
\hline
d=2
& \begin{tabular}{c}
    Single pass, 10mm PPLN \\
    (1475nm+803nm=527nm)\\
    \end{tabular}
&\begin{tabular}{c}
   Coherent laser:\\
   $\ket{L}+e^{i\phi}\ket{-L}$ \\
 \end{tabular}
&  \begin{tabular}{c}
 1.8@L=0\\
1.3@L=1\\
    \end{tabular}
&\cite{steinlechner2016frequency}\\
\hline
d=2
& \begin{tabular}{c}
    Single pass, 10mm PPLN \\
    (1565nm+806nm=532nm)\\
    \end{tabular}
&\begin{tabular}{c}
   Coherent laser of LG and HG mode
 \end{tabular}
&  Average CE$~ 10^{-4}$
&\cite{sephton2019spatial}\\
\hline
d=3,5
& \begin{tabular}{c}
    Single pass, 10mm PPKTP \\
    (1550nm+794nm=525nm)\\
    \end{tabular}
&\begin{tabular}{c}
   Single photon qudit state:\\
   $\ket{-2}+\ket{-1}+..+\ket{2}$ \\
 \end{tabular}
&  \begin{tabular}{c}
 0.37@L=0\\
0.42@L=1\\
0.33@L=2\\
0.24@L=3\\
    \end{tabular}
&This work\\
\hline
\end{tabular}
\caption{Conversion efficiency of various OAM states.}\label{T1}
\end{table*}

\section{Quantum state tomography of a qudit}
For a qudit ${\left| \psi  \right\rangle _d}{\rm{ = }}\frac{1}{{\sqrt d }}\sum\nolimits_{j =  - [d/2]}^{[d/2]} {\left| j \right\rangle } $, the density matrix can be written as:
 \begin{equation}\label{6}
   \rho {\rm{ = }}{\left| \psi  \right\rangle _d} \otimes {\left\langle \psi  \right|_d}
 \end{equation}
Where each of OAM eigenstate should be expressed as a vector in a d-dimensional space, i.e., $\left| { - 2} \right\rangle {\rm{ = }}{\left[ {1\;0\;0\;0\;0} \right]^T}$, $\left| { - 1} \right\rangle {\rm{ = }}{\left[ {0\;1\;0\;0\;0} \right]^T}$, ..., $\left| 2 \right\rangle {\rm{ = }}{\left[ {0\;0\;0\;0\;1} \right]^T}$ in a five-dimensional space. For a qudit state, using this definition, we can calculate the theoretical matrix of a qudit \cite{thew2002qudit}. The density is a square matrix that each value is $1/d$. Experimentally, the density matrix of a qudit can be reconstructed by using high-dimensional quantum state tomography (QST) via mutually unbiased bases (MUBs):
\begin{equation}\label{7}
  \rho {\rm{ = }}\frac{1}{d}N\sum\limits_{i,j{\rm{ = }}0}^{{d^2} - 1} {{{\left( {A_i^j} \right)}^{ - 1}}{n_i}{{\hat \lambda }_j}}
\end{equation}
Where $\hat{\lambda}_j$ represents the elementary matrix associating SU(d) group; $n_j$ is the coincidence count between signal and heralding photon; $A_i^j = \left\langle {{a_i}} \right|{\hat \lambda _j}\left| {{a_i}} \right\rangle $ is the  measurement matrix associating with MUBs, $\left| {{a_i}} \right\rangle $.

The group of mutually unbiased bases (MUBs) $\left\{ {\left| {a_m^j} \right\rangle } \right\}$ can be generated using the Weyl group, Hadamard matrix, and Fourier Gauss transform methods. Here, we used the discrete Fourier Gauss transform to product MUBs in prime dimensional space \cite{wiesniak2011entanglement},
\begin{equation}\label{8}
  \left\{ {{\rm{|a}}_m^j{\rm{ > }}} \right\} = \left\{ {\frac{1}{{\sqrt d }}\sum\limits_{n = 0}^{d - 1} {\omega _d^{\left( {j{n^2} + nm} \right)}} \left| n \right\rangle } \right\}
\end{equation}
Where $j(j=0...d-1)$ indexes the group of the MUBs; $m(m=0...d-1)$ indexes the superposed OAM states for each set in MUBs, and ${\left| {\left\langle {a_m^j} \right|\left. {a_{m'}^{j'}} \right\rangle } \right|^2} = 1/d\left( {1 - {\delta _{jj'}}} \right)$ for the MUBs. In actuality, $j$ runs from 0 to d, with the last set of MUBs being the OAM eigenstates.
\begin{equation}\label{9}
  \left\{ {\left| {a_m^d} \right\rangle } \right\} \to \left\{ {|0 > ,|1 > ,\;...\;|d - 1 > } \right\}
\end{equation}
For a two-dimensional OAM space (d=2), the eigenstates of three Pauli operators form a complete set of MUBs, which can be represented by the following Pauli matrices. The corresponding measured superposed OAM states are:
\begin{equation}\label{10}
  \begin{array}{l}
{\rm{\{ }}{{\rm{I}}_1}{\rm{\}  = }}\left\{ {|0 > ,|1 > } \right\}\\
{\rm{\{ }}{{\rm{I}}_2}{\rm{\}  = }}\left\{ {\frac{{|0 >  + |1 > }}{{\sqrt 2 }},\frac{{|0 >  - |1 > }}{{\sqrt 2 }}} \right\}\\
\{ {{\rm{I}}_3}\}  = \left\{ {\frac{{|0 >  + i|1 > }}{{\sqrt 2 }},\frac{{|0 >  - i|1 > }}{{\sqrt 2 }}} \right\}
\end{array}
\end{equation}

We can find that the number of MUBs is $d(d+1)$, while the elements of the density matrix are $d^2$. The reason is that these MUBs form an overcomplete tomography basis \cite{Giovannini2013}. In principle, the density matrix can be reconstructed by Eq. B2. However, the density matrix may not be a physical density matrix; i.e., it has the property of positive semi-definiteness. For overcoming this disadvantage, We employ the maximum likelihood estimation method is to estimate a physical density matrix \cite{James2001,{liu2018coherent}}.
\begin{equation}\label{11}
  L({t_1},{t_2},...,{t_{{d^2}}}) = \sum\limits_{j = 0}^{{d^2} - 1} {\frac{{{{[N({{\left\langle \Psi  \right|}_j}{\rho _{\exp }}{{\left| \Psi  \right\rangle }_i}) - {n_j}]}^2}}}{{2N({{\left\langle \Psi  \right|}_j}{\rho _{\exp }}{{\left| \Psi  \right\rangle }_i})}}}
\end{equation}
Where ${\left| \Psi  \right\rangle _j}$ has the same meaning as the formula in reference \cite{James2001}.

\bibliography{HD_QFC}

\end{document}